\documentstyle[times,pramana,epsf,floats]{ias}
\begin{document}
\mark{{Heavy Ion Theory}{R J Fries}}
\title{High Energy Nuclear Collisions: Theory Overview}

\author{R J Fries}
\address{Cyclotron Institute and Department of Physics and Astronomy,
Texas A\&M University, College Station TX 77845, USA \\
and RIKEN/BNL Research Center, Brookhaven National Laboratory, Upton NY 11973,
USA}
\keywords{Heavy Ion Collisions, Quark Gluon Plasma}
\pacs{XX}
\abstract{
We review some basic concepts of Relativistic Heavy Ion Physics and discuss
our understanding of some key results from the experimental program
at the Relativistic Heavy Ion Collider (RHIC). We focus in particular
on the early time dynamics of nuclear collisions, some result from lattice 
QCD, hard probes and photons.}

\maketitle

\section{Introduction}

The universe a few microseconds after the Big Bang was filled with a 
hot and dense phase of matter. We believe that quarks and gluon at those
temperatures, above $10^{12}$ K, were deconfined and existed as a quark
gluon plasma (QGP). These ideas can be tested in collisions of nuclei
at ultra-relativistic energies. At the Relativistic Heavy Ion Collider
(RHIC) nuclei as heavy as gold are accelerated to an energy of 100 GeV
per nucleon. 40 TeV total energy is available in the collision of two of these
nuclei. A large fraction of this energy is used to create new particles 
(a few thousand of them) and to give them kinetic energies. 
The energy in the collision is
spread over a certain volume with peak energy densities of $\sim 10$-$30$ 
GeV/fm$^3$, much larger than in ordinary nuclear matter 
\cite{Adcox:2004mh,Adams:2005dq,Kolb:2003dz}. This process happens over
a very short time (a few fm/$c$) but we have strong indications from
observables like elliptic flow $v_2$ that
despite the short life times thermalization of the matter
is reached. The peak energy densities are safely above
the predicted critical energy densities $\approx 1$ GeV/fm$^3$ for a 
phase transition from hadrons to a quark gluon plasma 
\cite{Cheng:2007jq}.

In the first decade of RHIC a large amount of experimental results 
has been gathered. Some results have confirmed expectations, like
jet quenching. Others have turned out to be more diffult to decipher, 
e.g.\ signals from quarkonia that promised to be simple plasma thermometers
\cite{Mocsy:2005qw}, 
yet others have been great surprises, like the indications for 
a perfect fluid created at RHIC. We do have firm evidence that 
we create equilibrated quark-gluon matter in the experiments at RHIC.
\cite{Adcox:2004mh,Adams:2005dq,Gyulassy:2004zy}

In order to describe nuclear collisions at high energies and to filter
out the properties of quark gluon plasma from the stream of experimental
data we can slice the data in a simple way. Thermalization
of particles does not extend beyond 2 GeV/$c$ in transverse momentum $P_T$ 
around midrapidities, i.e.\ around 90$^\circ$ scattering angle from the 
beam axis.
However, this includes 99\% of all particles produced, the bulk of the
collision. The tail of the hadron spectrum at high $P_T$ does not come
from thermalized matter but from QCD jets, yet it bears the imprint of the quark
gluon plasma that the particles had to traverse in order to arrive at 
the detectors.
It had already been realized in the 1980s that we can use high-$P_T$ hadrons
as ``hard probes'' of the quark gluon plasma phase \cite{bjorken}.

In this overview we briefly cover some of the most fundamental approaches
to heavy ion collisions from the theoretical side. We will discuss 
initial interactions and the color glass condensate, the current state
of the lattice QCD results, jet quenching, and photons as an example
for electromagnetic probes. We have to forgo some other very timely
and important topics like heavy quarks, dileptons or hydrodynamics. 
The latter will be covered in an accompanying article by R.\ Bhalerao
\cite{Bhalerao:2010wf}.

\section{First Interactions}

Interactions of hadrons or nuclei at small momentum transfers, below
1-2 GeV/$c$ can not be described by standard perturbative QCD (pQCD) 
techniques. pQCD assumes rather dilute parton densities in the colliding
hadrons and is valid at weak coupling. Experimental data show that the gluon
distribution grows dramatically for larger energies (or small parton
momentum fraction $x$) which must eventually lead to a breakdown
of the standard pQCD picture. At very large energies the 
distribution of gluons in hadrons (and even more so in large nuclei) has 
to saturate, with gluon splittings and recombination happening 
in detailed balance. It has first been argued by McLerran and
Venugopalan that in such a saturation scenario the gluon density 
$\sim Q_s^{-2}$ sets a new scale, the
saturation scale, and that an effective theory can be constructed based
on a quasi-classical approximation to QCD 
\cite{McLerran:1993ka,McLerran:1993ni,Kovner:1995ja}. 
For $Q_s \gg \Lambda_{\mathrm QCD}$ 
the coupling can still be weak, although the theory has non-trivial
many-body effects included. This effective theory of QCD at
very high collision energies, called the color glass condensate, describes
a nuclear collision as the collision of two (highly Lorentz-contracted)
sheets of color $SU(3)$-charges \cite{Gelis:2010nm,Lappi:2010ek}.

\begin{figure}[htbp]
\epsfxsize=8cm
\centerline{\epsfbox{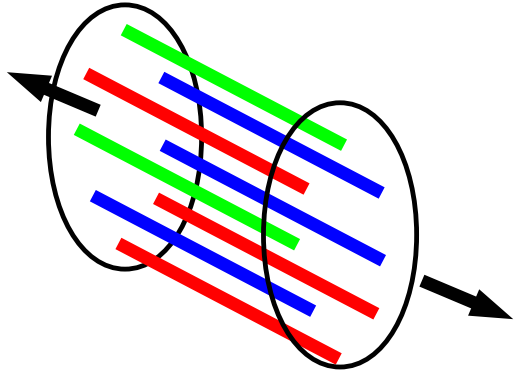}}
\caption{Two Lorentz-contracted nuclei shortly after the collision. The 
large momentum partons have gone through each other with small disturbance.
Their gluon fields have interacted to form strong longitudinal fields 
which connect the receding nuclei and slow them down. The scenario is 
reminiscent of a ``color'' capacitor with moving plates.}
\label{fig:flux}
\end{figure}

The initial gluon fields $A_1^\mu$, $A_2^\mu$ in the two nuclei before the 
collision are dominated by their \emph{transverse} electric and magnetic 
components. After the collision, this creates extremely strong 
\emph{longitudinal} (color-)electric and magnetic fields
\cite{Fries:2006pv,Lappi:2006fp}
\begin{equation}
  E_0 = ig[A_1^i,A_2^i] \, , \qquad B_0 = ig \epsilon^{ij} [A_1^i,A_2^j]
\end{equation}
(expressed here in a suitable axial gauge), see Fig.\ \ref{fig:flux}.
Note that the nuclei themselves (represented by the conserved
net baryon number carried by the valence quarks) go through each other.
Valence quarks (and other large-$x$ partons) only scattering 
occasionally to create jets. The longitudinal fields expand in the
space between the nuclei as they recede from each other. We now understand
that these strong longitudinal fields are the means by which energy density is 
deposited in the center of the colliding system. The initial energy 
density is given by
\begin{equation}
  \epsilon_0 = \frac{1}{2} \left( E_0^2 + B_0^2 \right) \approx
  \frac{g^6 N_c(N_c^2-1)}{8\pi} \mu_1^2 \mu_2^2 \ln^2 \frac{Q^2}{\Lambda^2} 
\end{equation}
where the $\mu_i^2$ are the densities of color charges in the two colliding
nuclei and $Q_0$ and $\Lambda$ are ultraviolet and infrared cutoffs.

The initial energy momentum tensor is diagonal with maximum pressure
anisotropy. The ``transverse pressure'' --- a slight abuse of terms so 
far from thermal equilibrium --- is $p_T = T^{xx} = T^{yy} = \epsilon_0$ and 
the ``longitudinal pressure'' is $p_L = T^{zz} = -p_T = -\epsilon_0$.
The negative longitudinal pressure corresponds to the fact that the
fields slow down the sources and hence the nuclei themselves (similar to the 
electric field attracting two capacitor plates in electrodynamics). 
The strong longitudinal fields lead to the creation of transverse fields
as time progresses and the anisotropy between transverse and longitudinal
components of the energy momentum tensor decreases. Equilibration 
needs two important ingredients: the gluon field has to decohere into gluons 
and quark-antiquark pairs in order to make the longitudinal pressure positive,
and eventually the pressure has to isotropize completely. The
exact way in which thermalization into an equilibrated quark gluon plasma 
proceeds, at least within the very short time (< 1 fm/$c$) suggested by 
data, is unknown 
\cite{Baier:2000sb,Xu:2007aa}. Plasma instabilities have recently been
suggested as a possible mechanism \cite{Mrowczynski:2005ki,Dumitru:2006pz}. 
The time for decoherence, a necessary ingredient for thermalization, 
was recently estimated to be proportional to $1/Q_s$ with a coefficient 
of order 1 \cite{Fries:2008vp,Fries:2009wh}.
For RHIC energies where $Q_s$ is estimated to be in the range
1-2 GeV, decoherence times are thus consistent with the experimentally 
observed thermalization times. Once thermalization is reached, the expansion
and cooling of quark gluon plasma is very well described by hydrodynamics.
The strong longitudinal gluon fields could play an important role in
two recent discoveries at RHIC: long-range rapidity correlations which
could be a direct image of the elongated flux tubes \cite{Gavin:2008ev}, 
and parity violation from quarks interacting with topologically 
non-trivial gluon configurations associated with 
$\mathbf{E}\cdot\mathbf{B} \ne 0$ 
\cite{Kharzeev:2007jp}.

\section{Lattice Estimates}

In lattice QCD the partition function of QCD is evaluated on a discretized 
Euclidean space-time grid using Monte-Carlo techniques. 
Results have long confirmed our expectation, originally from considerations 
of asymptotic freedom, that quarks and gluons are deconfined at high 
temperatures and that chiral symmetry is restored.
There are strong indications from realistic simulations of QCD with two 
light and one heavier quark flavors (for $u$, $d$, and $s$ quarks resp.)
that the phase transition is not of first order at vanishing net baryon 
densities, but a smooth cross over around a critical 
temperature $T_c$. Hence  we do not expect a sharp transition, say of 
the energy density, but rather the thermodynamic properties change
rapidly but smoothly as a function of temperature around $T_c$.

\begin{figure}[htbp]
\epsfxsize=8cm
\centerline{\epsfbox{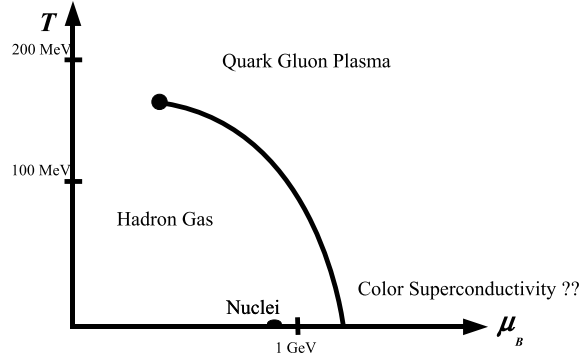}}
\caption{Schematic phase diagram of QCD. The first-order phase transition
line ends in a critical point, at smaller baryon densities the transition
between hadronic matter and quark gluon plasma is a cross over.}
\label{fig:phases}
\end{figure}

Lattice QCD at finite baryon chemical potential $\mu_B$
suffers from an unpleasant sign problem that makes Monte-Carlo techniques
rather ineffective. In recent years, several techniques have been devised
(reweighting, Taylor expansion around $\mu_B=0$, etc.) to explore the
phase diagram of QCD away from zero net baryon density. Most calculations
agree that there is a critical point in a range $\mu_B \approx 200 \ldots
400$ MeV from which on the cross over line between hadronic and
quark gluon matter becomes a first order phase transition.  
Our understanding of the QCD phase diagram is shown schematically
in Fig.\ \ref{fig:phases}.

Many modern calculations use close to realistic light quark masses with
pion masses as small as 200 MeV. Estimates for the critical temperature 
at small baryon densities --- the domain in which RHIC at top 
energies and the Large Hadron Collider (LHC) are operating --- range from 
176 to 192 MeV \cite{Aoki:2006br,Cheng:2006qk}. This is a rather
large discrepancy for results from different groups. While
many results hint at one common temperature for the chiral and deconfinement 
phase transition, one group sees evidence for a chiral 
transition at 151 MeV while they measure the deconfinement transition at 
176 MeV \cite{Aoki:2006br}. The divergence of predictions has 
recently attracted a lot of attention and is under investigation.

There is however consensus on the gross features of the QCD equation of state
calculated on the lattice. Around the critical temperature the normalized
energy density $\epsilon/T^4$ exhibits a large jump due to the drastic
increase in degrees of freedom going from the hadronic phase (basically
pions below $T_c$) to the deconfined phase with quarks and gluons.
$\epsilon/T^4$ and the reduced pressure $p/T^4$ do not reach their 
Stefan-Boltzmann limit even around $4 \times T_c$ which has been interpreted
as a hint for the strongly interacting nature of quark gluon plasma even
far above the phase transition \cite{Cheng:2007jq}. However, refined 
perturbative calculations can describe the lattice results
reasonably well down to about $2\times T_c$ \cite{Laine:2006cp}.

More recently it has been attempted to extract transport coefficients
like the shear viscosity $\eta$ and the bulk viscosity $\beta$ of quark
gluon plasma from lattice QCD calculations. As improved viscous 
relativistic hydrodynamic calculations become more and more available 
the interest in these new observables will increase \cite{Bhalerao:2010wf}.

\section{Hard Probes}

Jet quenching, the loss of high momentum particles in nuclear collisions
had long been predicted theoretically \cite{bjorken}. When it was 
finally observed in 
early RHIC data the effect turned out to be very large. Nearly 80\% of the
particles expected at a given transverse momentum $P_T$ were suppressed in
central collisions of gold nuclei. We quantify this suppression most
conveniently by the nuclear modification factor 
\begin{equation}
  R_{AA} = \frac{dN^{AA}/P_T}{\langle N_{\mathrm{coll}} \rangle dN^{pp}/dP_T}
\end{equation}
which is the ratio of particle yields in nucleus-nucleus vs proton-proton
collisions modulo the number of binary collisions expected in a nuclear
collision. For a loose superposition of nucleon collision the modification
factor would be (close to) unity. In fact there are a few ``cold'' nuclear
matter effects which lead to deviations of $R_{AA}$ from unity. These include
modifications of parton distributions for nucleons bound in nuclei vs 
free nucleons (known as shadowing and the EMC effect), and multiple 
scattering in the initial state which leads to the so-called Cronin effect 
\cite{Cronin:1974zm}. At top RHIC energies we expect particles produced
with a few GeV energy around midrapidity to exhibit anti-shadowing, i.e.\
a slight enhancement, which gives way to a suppression at larger $P_T$.
The Cronin effect leads to an enhancement for the same particles at a
few GeV which dies out toward higher $P_T$.
However, all of these cold nuclear matter effects are small, and partially
competing with each other. In summary they lead to deviations of $R_{AA}$ 
from unity of less than 20\%.

Two effects are important to understand the energy loss of a fast
quark or gluon propagating through quark gluon plasma. First, the 
dominating mechanism for energy loss, at least for light quarks and gluons,
is through induced gluon radiation. This leads to larger suppression than
energy loss than from elastic collisions. Secondly, the dependence of the 
energy loss on
the path length $L$ is quadratic, $\Delta E \propto L^2$. This LPM effect
(after Landau, Pomeranchuk and Migdal who first described the similar
effect in quantum electrodynamics \cite{Landau:1953um,Migdal:1956tc}) 
comes from a destructive interference.
Consider a parton radiating a gluon with energy $\omega$ and relative
transverse momentum $k_T$. The formation time of the parton-gluon system
is given by $\tau_f=\omega/k_T^2$, and further radiation during this time is
suppressed. In other words, if the mean free path $\lambda$ of the parton 
is smaller than the typical formation time $\omega/k_T^2$, then the
scattering off $N_{\mathrm{coh}}=\tau_f/\lambda $ number of partons 
happens coherently. This leads to a differential energy loss
$dE/dx = -\hat q x$ per unit path length, and to the famous quadratic 
path dependence of the total energy loss.
The transport coefficient $\hat q$ parameterizes the momentum transfer 
between the medium and the parton. It is given by the momentum transfer
squared $\mu^2$ per mean free path,
\begin{equation}
  \hat q = \frac{\mu^2}{\lambda} \, .
\end{equation}
Comparisons of $\hat q$ extracted from measurements at RHIC and from
cold nuclear matter experiments, e.g.\ HERMES, show a large increase
of the quenching power at RHIC \cite{Wang:2002ri}.

The handwaving arguments above can be backed up by calculations
which compute energy loss under certain simplifying assumptions.
The most important models based on perturbative QCD are
\begin{itemize}
\item the Higher Twist (HT) formalism by Guo and Wang 
  \cite{Guo:2000nz,Wang:2001ifa,Wang:2002ri}. It uses results from a 
  rigorous calculation
  of medium-modified fragmentation functions in semi-inclusive 
  deep-inelastic electron scattering off nuclei ($e+$A) and transfers these
  fragmentation functions to nuclear collisions.
\item the AMY formalism based on a series of papers by Arnold, Moore and Yaffe
  \cite{Arnold:2002ja,Jeon:2003gi}. These authors use hard thermal loop 
  resummed perturbation theory, valid at asymptotically large temperatures,
  to calculate the rate of energy loss for fast partons
\item the ASW formalism developed by Armesto, Salgado and Wiedemann 
  \cite{Salgado:2002cd,Salgado:2003gb}. Here multiple soft gluon emission 
  is resummed using a Poisson statistics for the number of gluons emitted. 
\item The GLV approach by Gyulassy, Levai and Vitev 
  \cite{Gyulassy:1999zd,Gyulassy:2000fs,Gyulassy:2000er}.
  They discuss scatterings off static scattering centers in an opacity 
  expansion in the medium.
\end{itemize}

\begin{figure}[htbp]
\epsfxsize=4cm
\centerline{\epsfxsize=5cm \epsfbox{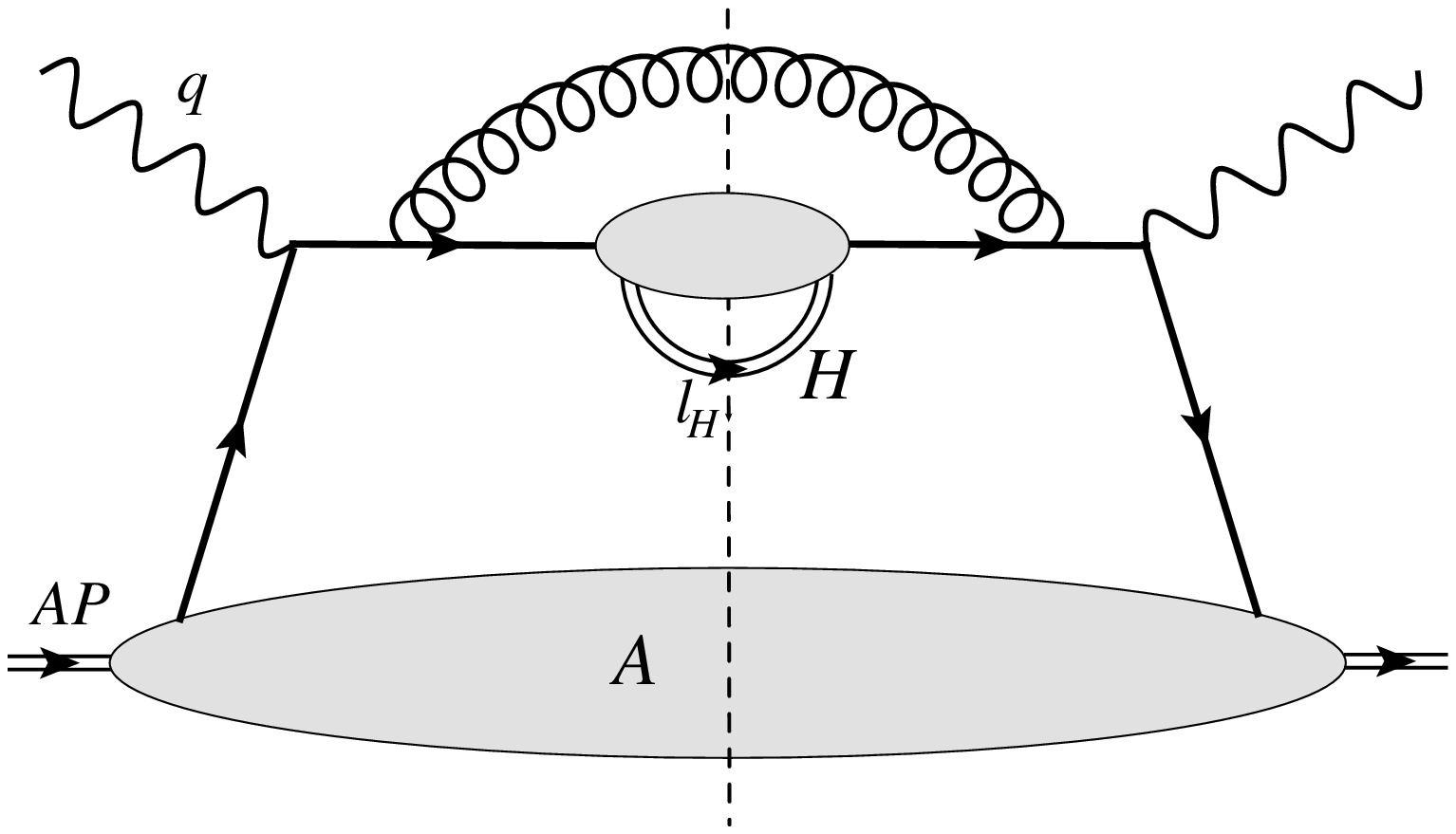} \qquad \epsfxsize=5cm
\epsfbox{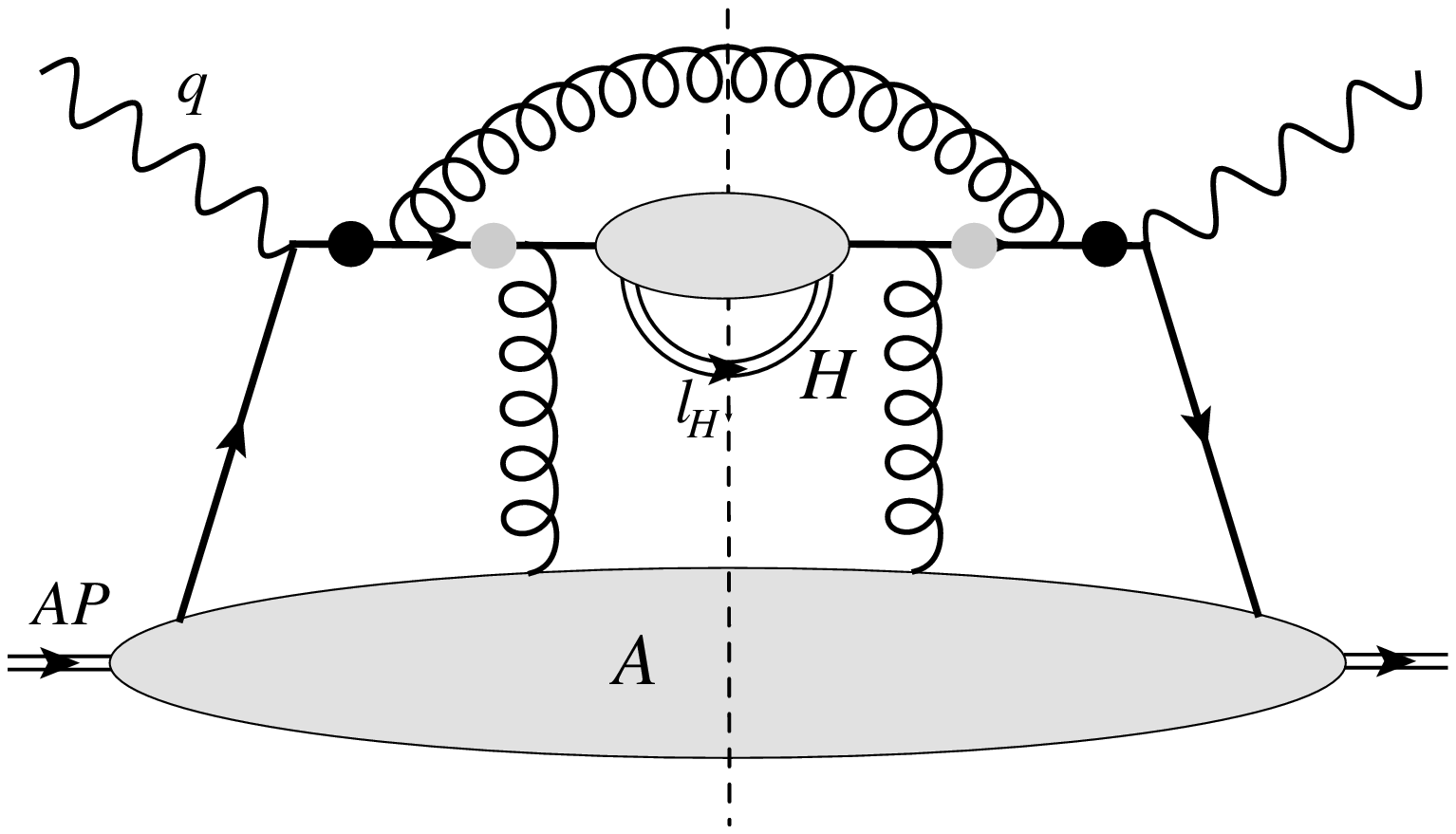}}
\caption{Left panel: diagram contributing to the DGLAP evolution equation
of (vacuum) fragmentation functions in semi-exclusive deep-inelastic
scattering. Right panel: typical diagram contributing to the modified
evolution equations at the twist-4 level. The outgoing parton scatters off
the medium and radiates.}
\label{fig:ht}
\end{figure}

As an example we briefly highlight some fundamental ideas of the Higher
Twist formalism.
Fig.\ \ref{fig:ht} shows typical diagrams contributing to the usual DGLAP
equations for fragmentation functions (left panel), and to a set of modified
evolution equations that are formally of higher twist in deep-inelastic
scattering (DIS). Higher twist
in factorized perturbation theory usually refers to sub-leading contributions
that are suppressed by powers of the ratio of the non-perturbative over 
the large perturbative scale, $\sim \Lambda^2/Q^2 << 1$. In nuclei, some
of those higher twist contributions are re-enhanced through a size factor,
such that they enter at a level $L \Lambda^3/Q^2 \sim 1$. Note that we
have indicated two propagators each in amplitude and complex conjugated
amplitude in the right panel through grey and black circles. The interference 
between the poles of two of those propagators leads to the LPM suppression 
effect in the Higher Twist formalism. The modified evolution 
equations 
define medium-modified fragmentation functions $\tilde D(z)$ which 
describe the energy loss of partons in semi-exclusive DIS.
In their simplest form they can be parameterized as
\begin{equation}
  \tilde D(z,Q) = \frac{1}{1-\Delta z} D\left(\frac{z}{1-\Delta z}\right) \, .
\end{equation}
where $z$ is the momentum fraction of the hadron within the fragmenting parton
and $\Delta z$ is the average shift due to induced radiation.
Similar fragmentation functions, with the average energy loss scaled up to fit
the data, can then be used to describe high-$P_T$ data at RHIC.

The energy loss models described above differ in some of their underlying
assumptions and are by no means completely equivalent. Nevertheless,
all of them manage to describe both the $P_T$-dependence and the impact
parameter-dependence of $R_{AA}$ measured at RHIC remarkably well after
\emph{one} parameter, typically $\hat q$ or a related quantity, is fixed.
With the advent of data on hadron correlations at high $P_T$ additional 
non-trivial restrictions on fits were available. Typically experiments
measure pairs of hadrons, a trigger hadron $t$ and an associated hadron 
$a$ which are then presented as a per-trigger yield
\begin{equation}
  Y(P_t,P_a,\Delta\phi) = \frac{dN/dP_t dP_a d(\Delta \phi)}{dN/dP_t}
\end{equation}
as a function of the transverse momenta of the trigger and associate
particle, $P_t$ and $P_a$ resp., and the relative azimuthal angle
$\Delta \phi$. To determine the suppression relative to free proton-proton
collisions we can once more look at the appropriate nuclear 
modification factor
\begin{equation}
  I_{AA} = \frac{Y^{AA}(P_t,P_a,\Delta\phi)}{Y^{pp}(P_t,P_a,\Delta\phi)} \, .
\end{equation}
$I_{AA}$ poses the first true challenge for jet quenching models, and
some studies have found incompatible values of $\hat q$ from separate
fits to $R_{AA}$ and $I_{AA}$. It has also become clear that
there is a long list of uncertainties, e.g.\ details of the modeling of the 
background fireball \cite{RFR}, the treatment of energy loss at early times before the 
formation of equilibrated quark
gluon plasma, etc.\ \cite{Armesto:2009zi}.

Where do we stand with hard probes? A comparative study by Bass et al.\
\cite{Bass:2008rv} found unacceptably large discrepancies between values 
of $\hat q$ derived from different jet quenching models fitted to RHIC data. 
The extreme values found in this study are
\begin{equation}
  \hat q = 18.5\>  \text{GeV}^2/\text{fm for ASW}\, , \qquad
  \hat q = 4.5 \> \text{GeV}^2/\text{fm for HT}\, .
\end{equation}
The good news is that these values are, by far, larger than for cold 
nuclear matter. However, they are not compatible among each other.
A sustained effort is now underway, spear-headed by the TECHQM initiative
and the JET collaboration, to systematically investigate hard probes,
by comparing and vetting different calculations, and by identifying and 
narrowing down the sources of theoretical uncertainties.

Once a reliable value of $\hat q$ is established, a comparison with
its perturbative value in an equilibrated plasma $\hat q_{\mathrm{pert}} 
= 2 \epsilon^{3/4}$ will tell us whether the coupling of jets
of the quark gluon plasma is weak or strong.
Some other interesting developments are the emergence of
simulations for the full evolution of jet showers in a medium 
\cite{Armesto:2009fj,Zapp:2008gi}, medium modifications to jet shapes 
\cite{Vitev:2008rz}, and hadron chemistry at large momentum $P_T$ 
\cite{Liu:2008zb}.

\section{Photons}

Hadrons from the bulk carry information from the point of
their last interaction (the freeze-out), and hadrons at large momentum $P_T$
carry some information of the fireball integrated over its history. 
On the other hand electromagnetic probes, i.e.\ real and virtual photons 
(dileptons) have the wonderful property that they do not re-interact once 
created, due to their large mean-free path in quark gluon plasma. 
Hence they are unique
penetrating probes that can give us unobstructed information about
the center of the fireball in nuclear collisions, and about
the earliest times during the collision. The following sources of photons
are important in relativistic heavy ion collisions \cite{Gale:2004ud}.
\begin{itemize}
\item Prompt hard photons from hard collisions: they are created from
  annihilation and Compton scatterings, $q+\bar q\to \gamma+g$ and 
  $q+g\to \gamma+q$ resp., of large momentum partons in the original
  nuclear wave functions. They dominate the direct photon spectrum at the 
  largest transverse momenta
  $P_T$. They also do not carry information about the quark 
  gluon plasma formed, but exhibit some of the initial state effects
  (shadowing, Cronin effect) mentioned above.
\item Fragmentation photons from vacuum bremsstrahlung: any initial hard 
  scattering of partons can lead to bremsstrahlung. Most of the time
  this will be gluon radiation, but there is a sizable contribution of 
  photons which can be described by parton-to-photon fragmentation functions.
  Vacuum fragmentation photons have a somewhat steeper $P_T$-spectrum
  but are comparable in yield with prompt hard photons except for the
  largest momenta. Their yield can be modified in nuclear collisions
  as partons travel over some distance and lose energy 
  before radiating a photon.
\item Photons from jet conversions and induced bremsstrahlung \cite{Fries:2002kt,Fries:2005zh,Zakharov:1996fv,Majumder:2007ne}: as leading
  jet partons travel through the medium and lose energy through induced
  gluon radiation, again part of this induced radiation can be in the form
  of photons. 
  In addition,
  elastic annihilation and Compton scatterings of fast jet partons with
  partons in the quark gluon plasma can lead to an effective conversion
  in which all of the momentum of the fast parton can be transferred to 
  the photon. These processes give competitive photon yields in 
  heavy ion collisions at RHIC and LHC at intermediate $P_T$
  of a few GeV/$c$. The photons carry the full 
  time evolution of the quark gluon plasma fireball as the local conversion
  rates are proportional to $T^2 \ln T$ and thus one can estimate
  plasma temperatures $T$ from their yields. 
\item Thermal photons from both the QGP and the hadronic phase 
  \cite{Kapusta:1991qp,Arnold:2001ba}: annihilation
  and Compton as well as bremsstrahlung processes between quarks and gluons 
  also play out in the equilibrated plasma, providing an exponential photon 
  spectrum that dominates at low $P_T$. Its slope obviously provides the
  temperature $T$, although the photons seen in the detector are integrated
  over the spatial profile as well as the time evolution of the fireball.
  The hot hadronic gas phase below the phase transition temperature also
  leads to thermal photons at lower temperatures.
\end{itemize}

All of these sources have been computed in a sustained effort over two 
decades. A modern, comprehensive calculation and comparison with data
can be found in publications of the McGill group \cite{Qin:2009bk}. They compare
well with the experimentally measured spectra and $R_{AA}$ of direct photons
(decay photons of hadrons like the $\pi^0$ bury the direct photon signal
in experimental data and have to be subtracted). At low $P_T$
there is ample evidence that the ``glow'' of the quark gluon plasma has
been found and that the temperatures are above the predicted values of 
$T_c$. Constraints on other sources, like the conversion photons, are not
as stringent. Other observables have been suggested to explicitly find those
photons, and to get a complementary picture of both the medium and 
jet quenching: e.g.\ the azimuthal asymmetry $v_2$ of photons, and
photon-hadron correlations. 

$v_2$ is the second order harmonic of the 
spectrum $dN/d\phi$ w.r.t.\ the azimuthal angle around the beam axis.
The reference axis ($\phi=0$) is given by the reaction plane. So far we have
not discussed $v_2$ for hadrons from the bulk fireball or for hadrons
from jets. In both cases $v_2$ is positive, meaning more particles are emitted
into the reaction plane than out of it. For bulk particles this comes from
the hydrodynamic expansion with larger pressure gradients and hence larger 
flow into the plane, for high-$P_T$ particles it derives from a smaller 
opacity and less absorption into the plane.
It has been predicted in \cite{Turbide:2005bz} that more conversion 
photons should be emitted in the elongated direction of the fireball, 
out of the reaction plane than into the reaction plane. This would lead 
to a negative $v_2$ for this photon source, the first process discovered 
to exhibit this behavior.
Its experimental discovery would unanimously confirm conversion and induced
bremsstrahlung of jets in medium. Since other photon sources come with 
vanishing or positive $v_2$ the sign of the total direct photon $v_2$ is 
rather unclear and the absolute value is small due to the cancellations.
Experimental data confirm small absolute values but the error bars in the
$P_T$ region of interest are too small to pin down the sign
\cite{Turbide:2005bz,Chatterjee:2008tp}.

One of the most exciting tools in our arsenal combine properties of 
hard and electromagnetic probes.These are photon-hadron correlations at 
high momentum. 
Photon-triggered
fragmentation functions can give unprecedented access to the
mechanisms of energy loss, as prompt hard photons carry the information
about the \emph{initial} momentum of the hard recoil quark or gluon.
This information is not available if $R_{AA}$ is considered and it is
considerably washed out even if $I_{AA}$ is considered. First experimental
results of photon-triggered fragmentation functions are available.

\section{Summary}

The processes unfolding in collisions of nuclei at high energies are
qualitatively understood. The system goes from a phase of strong initial
gluon fields through a thermalization process to a hydrodynamically
expanding quark gluon plasma which eventually hadronizes. QCD jets, 
high-$P_T$ hadrons and photons can serve as built-in probes of the new
phases of QCD. The field is now moving into a stage of quantitative analysis
in which transport coefficients and other observables should be extracted
with much reduced error bars. In the near future the high energy ion program
at LHC and the future FAIR facility will provide completely new perspectives
on the QCD phase diagram.

%\begin{table}[h]%\caption{hey what a nonsense.This is a long line checking if things
%work properly. may be . may be not . knock woods or keep fingers crossed? 
%How do you type keeping fingers crossed???}
%\hskip4pc\vbox{\columnwidth=26pc
%\begin{tabular}{lll}
%class & calory requirement & number of person \\ \hline
%army & 5000 & 1247 \\
%navy & 3895 & 2435 \\
%airforce & 3098 & 8274 \\ 
%\end{tabular}
%}
%\end{table}

\section*{Acknowledgments}

R.~J.~F. is supported by CAREER Award PHY-0847538 from the U.~S.~National
Science Foundation, RIKEN/BNL Research Center, and DOE grant DE-AC02-98CH10886.
He would like to express his gratitude to the organizers of ISNP 2009 for
their hospitality and kind invitation.

\end{document}